\documentstyle[12pt,amsfonts,amssymb,amsbsy,epsf]{article}
\newcommand{\Rsub}{\rm\scriptscriptstyle}
\begin{document}
\title{Renormalization group improvement of truncated 
perturbative series in QCD.\\ Decays of $\tau$-lepton and 
$\eta_c$-charmonium}
\author{V.V.Kiselev,\\[2mm]
\small Russian State Research Center ``Institute for
High Energy Physics'', \\ 
\small Division of Theoretical Physics, Protvino, 
Moscow Region, 142280 Russia\\
\small E-mail: kiselev@th1.ihep.su, Fax: -7-0967-744937}
\date{}

\maketitle
\begin{abstract}
We formulate a general scheme to improve the truncated perturbative 
expansion in $\alpha_s$ by means of the renormalization group in QCD 
for the single-scale quantities. The procedure is used for 
the evaluation of hadronic decay rates of $\tau$-lepton 
and $\eta_c$-charmonium. The scale dependence of result 
for $\eta_c$ is studied in the scheme of fixed value for 
the $\overline{\mbox{MS}}$-mass of charmed quark.
\end{abstract}

\section{Introduction}
\label{sec:1}

For many physical cases in QCD, an observable quantity is usually 
expressed in terms of truncated series in the coupling constant
$\alpha_s$ with given coefficients, so that in the 
next-to-next-to-next-to-leading order (N$^3$LO) we get
\begin{equation}
  \label{eq:1}
  {\cal R} = 1 + c_1\,\frac{\alpha_s(\Lambda)}{\pi}+
             c_2\,\left(\frac{\alpha_s(\Lambda)}{\pi}\right)^2+
             c_3\,\left(\frac{\alpha_s(\Lambda)}{\pi}\right)^3,
\end{equation}
where $c_{1\,,2,\,3}$ are some numbers, and $\Lambda$ is a fixed scale.
So, the value $\cal R$ is the single-scale quantity. The exhausted 
examples are the followings:
\begin{enumerate}
\item
The hadronic fraction of $\tau$-decay width \cite{1,PDG}
\begin{eqnarray}
&&  {\cal R}_{\tau}  = \frac{\Gamma[\tau\to \nu_{\tau}\mbox{hadrons}]}
  {\Gamma[\tau\to \nu_{\tau}e^+\nu_e]} = \nonumber\\ &&
{\cal R}_{\tau}^{[0]} \left(
   1 + c_1^{\tau}\,\frac{\alpha_s(m_{\tau})}{\pi}+
       c_2^{\tau}\,\left(\frac{\alpha_s(m_{\tau})}{\pi}\right)^2+
       c_3^{\tau}\,\left(\frac{\alpha_s(m_{\tau})}{\pi}\right)^3
  +\Delta r_{\mbox{\small NP}}\right),
  \label{eq:2}
\end{eqnarray}
where ${\cal R}_{\tau}^{[0]}=3.058$, the coefficients are given by
\begin{equation}
  \label{eq:3}
  c_1^{\tau}=1,\quad c_2^{\tau}=5.2,\quad c_3^{\tau}=26.4,
\end{equation}
and $\Delta r_{\mbox{\small NP}}=-0.014\pm 0.005$ is a nonperturbative 
contribution.
\item
The hadronic fraction of $\eta_c$-decay width \cite{2}
\begin{eqnarray}
  \label{eq:3}
  {\cal R}_{\eta_c} & = &\frac{\Gamma[\eta_c\to \mbox{hadrons}]}
  {\Gamma[\eta_c\to \gamma\gamma]} = 
{\cal R}_{\eta_c}^{[0]} \left(
   1 + d_1\,\frac{\alpha_s(2 m_c)}{\pi}\right),
\end{eqnarray}
where 
\begin{equation}
  \label{eq:4}
  {\cal R}_{\eta_c}^{[0]} = \frac{C_F}{2N_c}\,\frac{1}{e_c^4}\,
\frac{\alpha_s^2(2m_c)}{\alpha^2_{\rm em}}
\end{equation}
with $C_F=(N_c^2-1)/2 N_c$, $N_c=3$ is the number of colors, 
$e_c=2/3$ is the electric charge of charmed quark, and the coefficient
$d_1$ is given by
\begin{equation}
  \label{eq:5}
  d_1 = \frac{199}{6}-\frac{13\pi^2}{8}-\frac{8}{9}\,n_f-
\frac{2}{3}\ln 2,
\end{equation}
where $n_f=3$ is the number of `active' flavors, and $m_c$ is the pole
mass of charmed quark.
\end{enumerate}

The above formulae can be used for the extraction of $\alpha_s$ at the
appropriate scale. The value of $\alpha_s$-corrections is numerically
significant. So, the problem is how the truncated series can be improved.
The well-established approach to the solution of such the problem is 
a resummation of some significant terms. We mention two of such techniques.
The first is the summation of $(\beta_0\alpha_s)^n$ contributions, where
$\beta_0= 11 - \frac{2}{3}\, n_f$ is the first coefficient of 
$\beta$-function in QCD \cite{3,4}. 
The second procedure is based on an appropriate
change of renormalization scheme by $\bar\alpha_s=\alpha_s (1+b_1
\alpha_s+\ldots)$ to the given order in the coupling constant, which allows
one to decrease a role of higher-order corrections or even to minimize it
with the modification of $\bar \beta(\bar\alpha_s)$-function resulting 
in a different running of $\bar\alpha_s$ \cite{5}. The disadvantage of above 
methods is twofold. First, the next-order correction while computed exactly
can essentially differ from the approximation of $\beta_0\alpha_s$-dominance.
Second, the redefinition of renormalization scheme leads to the scale or
normalization-point dependence of matching procedure.

In this paper we present a procedure to improve the truncated series 
in the framework of renormalization group by introducing an auxiliary 
scale and taking a single-scale limit. A general formalism is given in 
Section \ref{sec:2}. The numerical estimates are presented in 
Section \ref{sec:3}. The analysis of scale dependence for the 
$\eta_c$-decay rate is performed, since the normalization 
at the pole mass involves the additional problem caused by the residual
change of $m_c$ by the variation of normalization point in the 
$\overline{\mbox{MS}}$-mass $\bar m_c(\mu)$ \cite{6}. 
Our results are summarized in Conclusion.

\section{Renormalization group improvement}
\label{sec:2}

For the sake of clarity, let us start with the consideration of 
first-order correction.
\begin{equation}
  \label{eq:6}
  {\cal K}={\cal R}/{\cal R}^{[0]} = 1+c_1 \frac{\alpha_s(\Lambda)}{\pi}.
\end{equation}
Introduce an auxiliary scale $\Lambda^\prime = \kappa\,\Lambda$, so that
\begin{equation}
  \label{eq:7}
  {\cal K}= 1+\frac{c_1}{\ln \kappa}\, \frac{\alpha_s(\Lambda)}{\pi}\,
\ln\kappa.
\end{equation}
Making use of the renormalization group relation to the first order 
in $\alpha_s$,
\begin{equation}
  \label{eq:8}
  \frac{\alpha_s(\Lambda)}{\alpha_s(\kappa\Lambda)} = 1+
\frac{\beta_0}{2\pi}\,\alpha_s(\Lambda)\ln\kappa,
\end{equation}
we clearly get
\begin{equation}
  \label{eq:9}
  {\cal K}(\kappa) = \left[\frac{\alpha_s(\Lambda)}{\alpha_s(\kappa\Lambda)}
\right]^{\displaystyle \frac{2 c_1}{\beta_0\ln\kappa}},
\end{equation}
which gives the ordinary presentation improved by the renormalization group.
Note, that one finds the limit
\begin{equation}
  \label{eq:9a}
  \lim_{\ln\kappa\to 0}\frac{\rm d}{{\rm d}\ln\kappa}\,{\cal K}(\kappa)
\equiv 0,
\end{equation}
which will be correct for the further consideration at a fixed order 
in $\alpha_s$.

The single-scale limit of $\ln\kappa\to 0$ can be easily evaluated
\begin{equation}
  \label{eq:10}
  {\cal K}^{\mbox{\sc rgi}}= \exp\left[c_1 \frac{\alpha_s(\Lambda)}{\pi}
\right],
\end{equation}
which is our result for the case of first-order correction.

In order to proceed with the higher-order corrections, let me perform 
the derivation in another way. So, the $\beta$-function  has the form
\begin{equation}
  \label{eq:11}
  \beta({\mathfrak a}) =  \frac{{\rm d}\ln {\mathfrak a}(\mu)}
{{\rm d}\ln \mu^2} = -\beta_0\,{\mathfrak a}-\beta_1\,{\mathfrak a}^2-
\beta_2\,{\mathfrak a}^3
\end{equation}
with ${\mathfrak a}=\frac{\alpha_s}{4\pi}$.
To the first order it gives
\begin{equation}
  \label{eq:12}
  \frac{\alpha_s(\Lambda)}{\alpha_s(\kappa\Lambda)} \approx 
\exp\left[\frac{\beta_0}{2\pi}\,\alpha_s(\Lambda)\ln\kappa\right],
\end{equation}
at $\ln\kappa\to 0$. Then,
\begin{equation}
  \label{eq:13}
  \left[\frac{\alpha_s(\Lambda)}{\alpha_s(\kappa\Lambda)}
\right]^{\displaystyle \frac{2 c_1}{\beta_0\ln\kappa}}\approx 
\exp\left[c_1 \frac{\alpha_s(\Lambda)}{\pi} \right],
\end{equation}
and expanding in $\alpha_s$, we rederive the renormalization group 
improvement (RGI) for the first-order correction.

Further,  we can easily find the RGI for the third order 
in $\alpha_s$ (N$^3$LO). Indeed, since
\begin{equation}
  \label{eq:14}
  \frac{\alpha_s(\Lambda)}{\alpha_s(\kappa\Lambda)} \approx 
\exp\left[(\beta_0+\beta_1\,{\mathfrak a}+
\beta_2\,{\mathfrak a}^2)\ln\kappa^2\right],
\end{equation}
we get
\begin{equation}
  \label{eq:15}
  \left[\frac{\alpha_s(\Lambda)}{\alpha_s(\kappa\Lambda)}
\right]^{\displaystyle \frac{c_1+4\bar c_2\,{\mathfrak a}+16
\bar c_3\,{\mathfrak a}^2}{\beta_0+\beta_1\,{\mathfrak a}+
\beta_2\,{\mathfrak a}^2}\frac{4}{\ln\kappa^2}} =  
\exp\left[c_1 \frac{\alpha_s(\Lambda)}{\pi}+\bar c_2 
\left(\frac{\alpha_s(\Lambda)}{\pi}\right)+\bar c_3 
\left(\frac{\alpha_s(\Lambda)}{\pi}\right) \right],
\end{equation}
where we put
\begin{equation}
  \label{eq:16}
  \bar c_2 = c_1-\frac{1}{2}\, c_1^2,\qquad
  \bar c_3 = c_3 -\frac{1}{6}\,c_1^3- c_1\,\bar c_2^2.
\end{equation}
Expanding in $\alpha_s$ at $\ln\kappa\to 0$, we find
$$
{\cal K}(\kappa) = \left[\frac{\alpha_s(\Lambda)}{\alpha_s(\kappa\Lambda)}
\right]^{\displaystyle \frac{c_1+4\bar c_2\,{\mathfrak a}+16
\bar c_3\,{\mathfrak a}^2}{\beta_0+\beta_1\,{\mathfrak a}+
\beta_2\,{\mathfrak a}^2}\frac{4}{\ln\kappa^2}}\approx
1 + c_1\,\frac{\alpha_s(\Lambda)}{\pi}+
             c_2\,\left(\frac{\alpha_s(\Lambda)}{\pi}\right)^2+
             c_3\,\left(\frac{\alpha_s(\Lambda)}{\pi}\right)^3.
$$
Thus, the third-order improved expression has the form
\begin{equation}
  \label{eq:17}
  {\cal K}^{\mbox{\sc rgi}}= \exp\left[ c_1\,\frac{\alpha_s(\Lambda)}{\pi}+
            \bar c_2\,\left(\frac{\alpha_s(\Lambda)}{\pi}\right)^2+
             \bar c_3\,\left(\frac{\alpha_s(\Lambda)}{\pi}\right)^3\right]
\end{equation}
We stress the renormalization group motivation used in contrast 
to {\em ad hoc} method of Pad\'e approximants. 

Let us show how the improvement works in a simple example. So, we consider a rather oscillating sum,
$$
{\cal E} = 1 -0.5+0.3 =0.8,
$$
which reveals a `slow' convergency, since
$$
{\cal E}^{[0]}=1,\quad 
{\cal E}^{[1]}=0.5,\quad
{\cal E}^{[1]}=0.8,
$$
while
$$
{\cal E}^{\mbox{\sc rgi}}= \exp[1-0.5+(0.3-0.5^3)]
$$
results in 
$$
{\cal E}^{\mbox{\sc rgi}}_{[0]} = 1,\quad
{\cal E}^{\mbox{\sc rgi}}_{[1]} = 0.61,\quad
{\cal E}^{\mbox{\sc rgi}}_{[2]} = 0.72,
$$
which is `more stable'.

Thus, we expect that ${\cal K}^{\mbox{\sc rgi}}$ possesses a more numerical
stability in the truncated series. Of course, if a series is essentially 
asymptotic, the improvement cannot cancel a `bad' convergency.

Next, we have to mention the numerical problem often appearing with 
the $\alpha_s$-corrections to the amplitudes and the amplitudes squared if
those corrections are significantly large. Indeed, the correction 
to the amplitude
$$
{\cal A}={\cal A}^{[0]}(1+c_1\alpha_s)
$$
should lead to 
$$
{\cal A}^2 =\left({\cal A}^{[0]}\right)^2(1+2\,c_1\alpha_s),
$$
so that the ratio
$$
(1+2\,c_1\alpha_s)/(1+\,c_1\alpha_s)^2
$$
numerically deviates from unit. The RGI has no such the problem, 
since the exponent does not involve the above mismatching.

Finally, we stress that the RGI does not present some kind of resummation 
of higher orders. In the resummation technique one certainly suggests 
a form of higher-order terms. In contrast, we give the exact expression 
produced by the renormalization group. At small $\alpha_s$ as dictated 
by the perturbative paradigm, the expression can be expanded till the
appropriate order. Thus, one could claim that the RGI procedure looks like
overflying the accuracy. To my opinion, one should use the RGI point 
as a central value of the calculated quantity, while the expansion truncated 
to the given order would indicate a systematic error of numerical estimate.

\section{Numerical estimates}
\label{sec:3}

\subsection{Hadronic fraction of $\tau$-lepton width}
\label{sec:3.1}

The RGI formula for the $\tau$-lepton decays into hadrons reads off
\begin{eqnarray}
&&  {\cal R}_{\tau}^{\mbox{\sc rgi}}  = 
{\cal R}_{\tau}^{[0]} \left\{\exp\left[
       c_1^{\tau}\,\frac{\alpha_s(m_{\tau})}{\pi}+
      \bar c_2^{\tau}\,\left(\frac{\alpha_s(m_{\tau})}{\pi}\right)^2+
      \bar c_3^{\tau}\,\left(\frac{\alpha_s(m_{\tau})}{\pi}\right)^3\right]
  +\Delta r_{\mbox{\small NP}}\right\},
  \label{eq:17a}  
\end{eqnarray}
where
\begin{equation}
  \label{eq:18}
  \bar c_2^{\tau}=4.7,\quad
  \bar c_3^{\tau}=22.53.
\end{equation}
Implementing
$$
{\cal R}_{\tau}^{\rm exp} = 3.635\pm 0.014,
$$
we find
\begin{equation}
  \label{eq:20}
  \alpha_s(m_{\tau}) = 0.333\pm 0.009,
\end{equation}
which results in 
\begin{equation}
  \label{eq:21}
  \alpha_s(m_{\rm Z}) = 0.119\pm 0.001,
\end{equation}
where we include the experimental uncertainty, only. For the sake 
of comparison, the PDG value extracted by the same measurement 
of $\tau$ rate reads off $\alpha_s(m_{\tau}) = 0.353\pm 0.007(\mbox{exp})
\pm 0.030(\mbox{th}),$ which respectively gives $\alpha_s(m_{\rm Z}) = 
0.121\pm 0.003$. We point out that the theoretical uncertainty 
in PDG is slightly overestimated, to our opinion, since the 
displacement of central value extracted in two ways equals 
$\Delta\alpha_s =0.02$.

Thus, the preferable value of coupling constant following from 
the $\tau$-lepton hadronic width is given by
\begin{equation}
  \label{eq:22}
  \alpha_s(m_{\rm Z}) = 0.119\pm 0.002
\end{equation}
with the central point closer to the `world average'.

\subsection{Hadronic width of $\boldsymbol \eta_c$-charmonium}
\label{sec:3.2}

The problem with the estimate of hadronic width of $\eta_c$-charmonium 
is twofold. First, the scale setting in the $\alpha_s$-correction is 
beyond the accuracy, since its variation contributes to $\alpha_s^2$. 
So, we should put the arbitrary scale by 
\begin{equation}
{\cal R}_{\eta_c}  = 
{\cal R}_{\eta_c}^{[0]} \left(
   1 + d_1\,\frac{\alpha_s(\mu)}{\pi}\right).
\label{eq:4a}
\end{equation}
The second point is the prescription for the pole mass of charmed quark.
In the perturbative QCD, the pole mass is strictly defined. 
The relation between the $\overline{\mbox{MS}}$-running mass $\bar m_c(\mu)$
and the pole mass is known to the $\alpha_s^3$-order \cite{7}. 
Explicitly, to the $\alpha_s^2$-terms \cite{8} we put
\begin{equation}
m^{\rm pole} = \bar m(\mu) \left(1+ c_1(\mu) \frac{\alpha_s^{\overline{\Rsub
MS}}(\mu^2)}{4 \pi} +c_2(\mu) \left(\frac{\alpha_s^{\overline{\Rsub
MS}}(\mu^2)}{4 \pi}\right)^2\right),
\label{pole}
\end{equation}
with
\begin{eqnarray}
c_1(\mu) &=&  C_F (4+3 {L}),\\
c_2(\mu) &=&  C_F C_A \left(\frac{1111}{24}- 8 \zeta(2)- 4 I_3(1)+\frac{185}{6}
L+\frac{11}{2}L^2\right)\nonumber\\
&&- C_F T_F n_f \left(\frac{71}{6}+ 8 \zeta(2) + \frac{26}{3} L + 2 L^2
\right)\\
&& + C_F^2 \left(\frac{121}{8}+30 \zeta(2)+8 I_3(1)+\frac{27}{2} L +
\frac{9}{2}L^2\right) - 12 C_F T_F (1-2 \zeta(2)),\nonumber
\end{eqnarray}
where $I_3(1) = \frac{3}{2}\zeta(3)- 6 \zeta(2)\ln 2$, and $L=2\ln(\mu/m^{\rm
pole})$.
Th evalue of pole mass is the renormalization invariant. However,
 at reasonable scales $\mu$, the residual dependence due to the truncation
of perturbative series is numerically significant. The reason of such 
the dependence is a growth of coefficients in series as caused 
by the renormalon. In fact, the pole mass becomes a scale-dependent quantity.
To avoid this problem, the operative procedure is to fix a short-distance
mass $m_{\sc qcd}$ free off the renormalon and to perform the calculations 
with the series expressed in terms of $m_{\sc qcd}$. We exploit two schemes, 
which lead to results close enough to each other.

\begin{figure}[ht]
  \begin{center}
\setlength{\unitlength}{1mm}
    \begin{picture}(100,70)
\put(3,3){\epsfxsize=100\unitlength \epsfbox{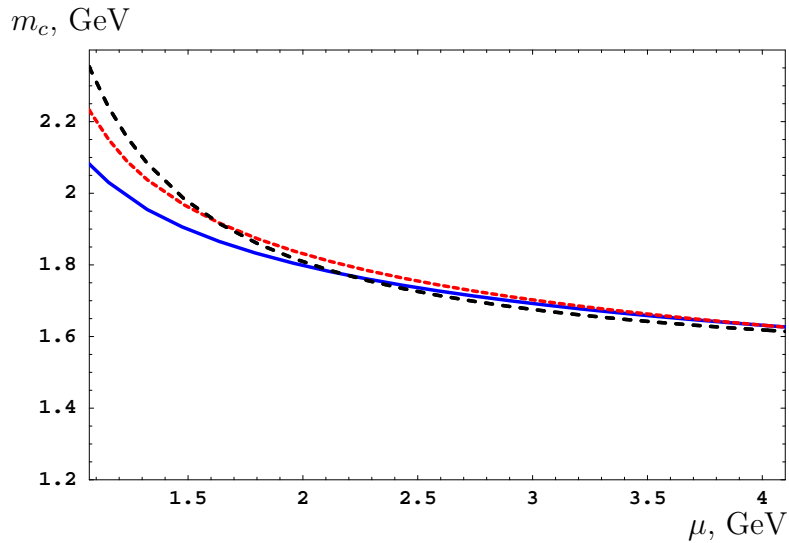}}
\put(90,0){$\mu$, GeV}
\put(0,67){$m_c$, GeV}
    \end{picture}

    \caption{The pole mass of charmed quark calculated in two schemes 
             versus the normalization scale. The dashed line gives the 
             result of matching in the potential approach, the solid line
             does by the perturbative relation between the pole and running
             masses shifted with $-\Delta m$ in (\ref{eq:23}) 
             at $\alpha_s(m_{\sc z})=0.118$, the short-dashed 
             line is the same as
             the solid one but at $\alpha_s(m_{\sc z})=0.121$. }
    \label{fig:1}
  \end{center}
\end{figure}

The first scheme is given by the $\overline{\mbox{MS}}$-running 
mass $\bar m(\mu)$. Taking 
$$
\bar m_c(\bar m_c) = 1.4\;\mbox{GeV},
$$
we calculate the pole mass shown in Fig. \ref{fig:1}. We have checked 
that the implication of RGI procedure to the relation between 
the pole and running masses is consistent with the above result, and 
the effect of RGI can be absorbed into the decrease 
of $\bar m_c(\bar m_c)$-value by about $50$ MeV, which below 
the systematic accuracy of matching procedure as discussed below. 

The second is the potential scheme described in ref \cite{KKO}. In this case, 
we calculate the scale-dependent matching of perturbative 2-loop scatic
potential $V_{{\rm pert}}(r,\mu)$ involving the 3-loop running $\alpha_s$
with the phenomenological QCD-motivated static potential $V(r)$ containing both
the 2-loop short-distance coulomb-like contribution as well as 
the long-distance linear confining term preserving the infrared stability. 
Then, the potential and, hence, the $V$-masses are free off the renormalon.
The heavy quark masses are fixed by the measured spin-average mass-spectra of
heavy quarkonia. So,
\begin{equation}
m_c^{\Rsub V} = 1.468\;{\rm GeV,}\quad
m_b^{\Rsub V} = 4.873\;{\rm GeV.}
\label{mcmb}
\end{equation}
The matching of scale-dependent perturbative potential $\delta V(\mu)=
V(r)-V_{{\rm pert}}(r,\mu)$ is extracted numerically as described 
in ref \cite{KKO}.
Thus, the cancellation of renormalon in the sum of $2 m^{\rm pole}+
V_{{\rm pert}}(r,\mu)$ gives
\begin{equation}
  \label{eq:23}
  m_c^{\rm pole}(\mu) = m_c^{\Rsub V}+\frac{1}{2}\delta V(\mu)-\Delta m,
\end{equation}
up to a constant shift $\Delta m$ independent of the scale. The matching with
the perturbative pole mass in (\ref{pole}) gives $\Delta m = -155\pm 15$ MeV,
depending on the variation of coupling constant $\alpha_s(m_{\sc z})$ 
in the limits of $0.118-0.123$. The value of $\Delta m$ indicates the accuracy
of matching procedure. The result is presented in Fig.\ref{fig:1}, 
which reveals a good agreement of two schemes used.

\begin{figure}[th]
  \begin{center}
\setlength{\unitlength}{1mm}
    \begin{picture}(100,70)
\put(3,3){\epsfxsize=100\unitlength \epsfbox{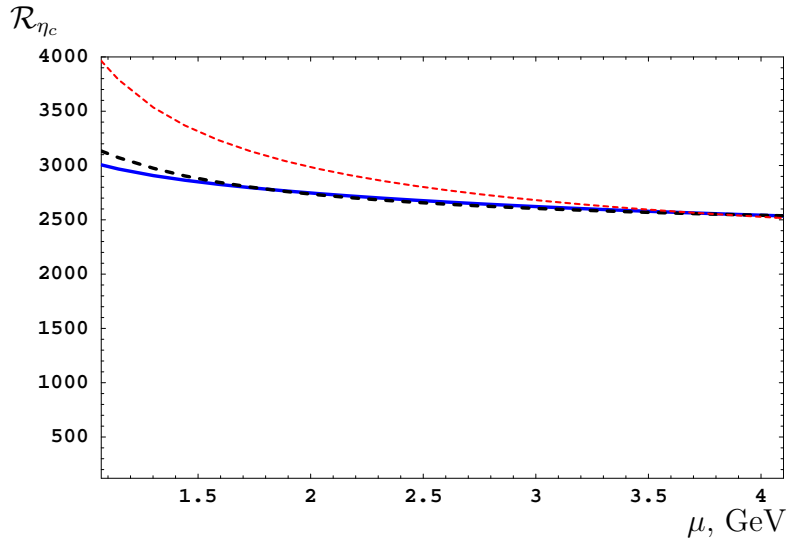}}
\put(90,0){$\mu$, GeV}
\put(0,67){${\cal R}_{\eta_c}$}
    \end{picture}
    \caption{The fraction of hadronic width for the $\eta_c$-charmonium 
     calculated with the fixed value of pole mass for the charmed quark 
     $m_c=1.64$ GeV (the short-dashed curve) and with the scale-dependent
     pole mass in the schemes of fixed running mass (the solid curve) and of
     potential approach (the dashed curve).}
    \label{fig:2}
  \end{center}
\end{figure}

Then, the perturbative formula (\ref{eq:4a}) with (\ref{eq:23}) results 
in the ${\cal R}_{\eta_c}$ shown in Fig. \ref{fig:2}, wherefrom we get
\begin{equation}
  \label{eq:24}
  {\cal R}_{\eta_c} = 2.6\cdot 10^3
\end{equation}
at $\mu = 3.9$ GeV with
$$
\alpha_s(2 m_c) = 0.242,\quad
\alpha_s(\mu) = 0.240,\quad
m_c = 1.64\;{\rm GeV}.
$$
The estimate in (\ref{eq:24}) is slightly greater than the value 
${\cal R}_{\eta_c} = 2.1\cdot 10^3$ given by Bodwin and Chen \cite{4}. 
We stress the scale-stability of our result.

Further, at the same scale we find
\begin{equation}
  \label{eq:25}
  {\cal R}_{\eta_c}^{\mbox{\sc rgi}} = 3.7\cdot 10^3.
\end{equation}
Then, comparing (\ref{eq:25}) with (\ref{eq:24}) we obtain the final estimate
including the theoretical uncertainty due to possible contributions 
of higher orders and, hence, the induced scale-dependence 
by the variation of central values as
\begin{equation}
  \label{eq:26}
  {\cal R}_{\eta_c}^{\rm th} = (3.7\pm 1.1)\cdot 10^3,
\end{equation}
which is in agreement with the experimental value
$$
  {\cal R}_{\eta_c}^{\rm exp} = (3.3\pm 1.3)\cdot 10^3,
$$
to be compared with ${\cal R}_{\eta_c}^{\sc nna} = 
(3.01\pm 0.030\pm 0.034)\cdot 10^3$ obtained in \cite{4} under 
the resummation of $(\beta_0\alpha_s)^n$-terms. We point out that 
the improvement of the experimental accuracy combined with the
calculation of $\alpha_s^2$-correction would give a good opportunity 
to extract the mass of charmed quark. In this respect, we refer to ref. 
\cite{mc}, where the $\alpha_s^2$-corrections were taken into account in 
the ratio of widths for the decays of $J/\psi \to e^+ e^-$ and 
$\eta_c \to \gamma \gamma$, so that the analysis suffers from the 
uncertainties related with the relativistic corrections entering the ratio 
for the different initial states. The advantage of ${\cal R}_{\eta_c}$ is the 
cancellation of such the initial state corrections.

\section{Conclusion}
\label{sec:4}

We have developed a general scheme to improve the estimate of 
truncated perturbative series in QCD by the tool of renormalization group 
for the single-scale quantities. The method allows one to get more 
realistic central values of the quantities as well as to estimate 
the theoretical uncertainty of results by comparison of RGI values with 
the perturbatively expanded ones. 
The RGI receipt for the calculation of quantity (\ref{eq:1}), (\ref{eq:6}) 
is given by (\ref{eq:16}) and (\ref{eq:17}).

We have applied the approach to the fractions of hadronic widths for the
$\tau$-lepton and $\eta_c$-charmonium, which allows us to get 
realistic estimates of $$\alpha_s(m_{\tau})\quad \mbox{and}\quad
{\cal R}_{\eta_c} = 
{\Gamma[\eta_c\to \mbox{hadrons}]}/{\Gamma[\eta_c\to \gamma\gamma]}$$
in a reasonable agreement with the appropriately measured values.

The author thanks prof.G.Bodwin for an exciting presentation of his results on 
the resummation technique for the hadronic fraction of $\eta_c$ width  
as he gave at the Heavy Quarkonium Workshop held in CERN, Nov. 8-11, 2002. 
A special gratitude goes to the organizing Committee of the Workshop, 
and personally to Antonio Vairo and Nora Brambilla for the invitation 
and a kind hospitality. I also thank prof.R.Dzhelyadin for the possibility 
to visit CERN in collaboration with the LHCB group, to which members 
I express my gratitude for a hospitatily. I thank prof.A.K.Likhoded, who 
asked me for the meaning of resummation technique, which initiated this work.

This work is in part supported by the Russian Foundation for Basic Research,
grant 01-02-99315.

\end{document}